# Van der Waals device integration beyond the limits of van der Waals forces via adhesive matrix transfer

Peter F. Satterthwaite[1], Weikun Zhu[2], Patricia Jastrzebska-Perfect[1], Melbourne Tang[1], Hongze Gao[3], Hikari Kitadai[3], Ang-Yu Lu[1], Qishuo Tan[3], Shin-Yi Tang[1,4,5], Yu-Lun Chueh[6,7], Chia-Nung Kuo[8,9], Chin Shan Lue[8,9,10], Jing Kong[1], Xi Ling[3,11,12], Farnaz Niroui[1*]

**Affiliations**
[1]Department of Electrical Engineering and Computer Science, Massachusetts Institute of Technology, Cambridge, MA 02139, USA
[2]Department of Chemical Engineering, Massachusetts Institute of Technology, Cambridge, MA 02139, USA.
[3]Department of Chemistry, Boston University, Boston, Massachusetts 02215, USA
[4]Department of Materials Science and Engineering, National Tsing Hua University, Hsinchu 30013, Taiwan
[5]Semiconductor Research Center, Hon Hai Research Institute, Taipei 11492, Taiwan
[6]Department of Materials Science and Engineering, National Tsing Hua University, Hsinchu 30013, Taiwan
[7]College of Semiconductor Research, National Tsing Hua University, Hsinchu 30013, Taiwan
[8]Department of Physics, National Cheng Kung University, Tainan 70101, Taiwan
[9]Taiwan Consortium of Emergent Crystalline Materials, National Science and Technology Council, Taipei 10601, Taiwan
[10]Program on Key Materials, Academy of Innovative Semiconductor and Sustainable Manufacturing, National Cheng Kung University, Tainan 70101, Taiwan
[11]Division of Materials Science and Engineering, Boston University, Boston, Massachusetts 02215, USA
[12]The Photonics Center, Boston University, Boston, Massachusetts 02215, USA

*Corresponding author. Email: fniroui@mit.edu

## Abstract

Pristine van der Waals (vdW) interfaces between two-dimensional (2D) and other materials are core to emerging optical and electronic devices. Their direct fabrication is, however, challenged as the vdW forces are weak and cannot be tuned to accommodate integration of arbitrary layers without solvents, sacrificial-layers or high-temperatures, steps that can introduce damage. To address these limitations, we introduce a single-step 2D material-to-device integration approach in which forces promoting transfer are decoupled from the vdW forces at the interface of interest. We use this adhesive matrix transfer to demonstrate conventionally-forbidden direct integration of diverse 2D materials ($MoS_2$, $WSe_2$, $PtS_2$, GaS) with dielectrics ($SiO_2$, $Al_2O_3$), and scalable, aligned heterostructure formation, both foundational to device development. We then demonstrate a single-step integration of monolayer-$MoS_2$ into arrays of transistors. With no exposure to polymers or solvents, clean interfaces and pristine surfaces are preserved, which can be further engineered to demonstrate both n- and p-type behavior. Beyond serving as a platform to probe the intrinsic properties of sensitive nanomaterials without the influence of processing steps, our technique allows efficient formation of unconventional device form-factors, with an example of flexible transistors demonstrated.

**Introduction**

Two-dimensional (2D) materials and their heterostructures present an emerging platform for novel optical and electronic devices[1]. Core to this potential is the ability to form van der Waals (vdW) heterostructures between 2D and other materials[2]. The universal nature of vdW interactions allows materials integration without the constraints of chemical and physical compatibility needed in conventional growth processes, such that the optical and electronic properties of the heterostructure can be tailored to achieve new and improved functionalities[3]. To form such heterostructures, the 2D material often needs to be transferred from its source to a target surface. Though universal, vdW interactions are weak and dependent on intrinsic materials properties[4]. Thus, they cannot be readily tuned at the interface of interest to promote arbitrary materials integration. Conventionally, this limitation is addressed using a sacrificial-layer[5], which supports the transfer, then, is removed through a combination of etching[6,7], burning[8,9], peeling[10,11] or melting[12,13]. This processing can introduce damage and contaminants[14], adversely affecting the desired properties. Furthermore, current platforms often require post-transfer fabrication, such as lithography, etching and deposition, to integrate the 2D material into functional devices, leading to device performance that can be limited by processing artefacts, rather than the heterostructure's intrinsic properties[15–19].

To avoid such adverse effects, an ideal approach would allow a single-step 2D material-to-device integration, without sacrificial-layers, solvents, high-temperatures, or post-transfer fabrication. Moreover, such dry transfer should be scalable and aligned, a feature not readily feasible with existing approaches. For a contact-and-release transfer to succeed, an intricate balance of surface interactions is critical – a higher adhesion is needed at the 2D-material/receiving substrate than at the 2D-material/source interface. Because vdW interactions cannot be tailored on demand, this requirement cannot be satisfied for arbitrary heterostructures, making vdW forces alone insufficient for vdW integration.

To address this fundamental limitation, we present adhesive matrix transfer. In this approach, the substrate of interest is embedded in a matrix with strong adhesive interactions to the 2D material, promoting its transfer. This decouples the properties of the functional interface, from forces required for its fabrication, enabling arbitrary pristine vdW heterostructures to form in a dry, sacrificial-layer-free and scalable manner, despite the small vdW interactions present. Using gold as an adhesive matrix, we demonstrate vdW integration of diverse two-dimensional materials with dielectric substrates, direct fabrication of which is conventionally forbidden. With a polymeric adhesive matrix and patterned graphene monolayers, we further show that our dry transfer can be aligned and scalable, here, presenting graphene-gold heterostructures and suspended 2D membranes. Lastly, we demonstrate a single step material-to-device integration by fabricating monolayer-$MoS_2$ transistor arrays on both rigid and flexible substrates with no exposure to solvents or sacrificial-layers, and no post-transfer fabrication. With surfaces and interfaces remaining pristine as a result, our approach allows for intrinsic materials properties to be studied, and for the device performance to be further tailored through surface engineering. In particular, utilizing charge-transfer doping, we demonstrate conversion of our intrinsically n-type MoS$_2$ transistor arrays into p-type devices with high on-off ratios.

**Results and Discussion**

**Solvent and polymer-free fabrication of pristine vdW heterostructures**

The semiconductor-dielectric heterostructure, such as monolayer-MoS$_2$ on SiO$_2$, is a core building block for optical and electronic devices. In an ideal case, this heterostructure would be fabricated by bringing a source MoS$_2$ crystal into conformal contact with SiO$_2$, and exfoliating a single monolayer over large areas without solvents or sacrificial-layers. The vdW interactions between these two surfaces are, however, weak. Thus, when attempted, no transfer is observed under optical microscopy or Raman mapping (Fig. 1a,d). In contrast, given the large vdW forces, large-area continuous monolayer MoS$_2$ can be transferred to ultrasmooth (< 1 nm root-mean-square roughness) template-stripped gold, as evidenced by the characteristic $E^1_{2g}$ and $A_{1g}$ Raman peaks[20] (Fig. 1b,e). Gold is, however, a limiting substrate as it is conductive. Previous approaches have addressed this using gold solely as a sacrificial-layer that is removed by wet-etching[21,22], or peeled after 2D layer transfer using an adhesive polymer surface[23], steps that can introduce damage and contaminants.

The adhesive matrix transfer approach allows the 2D material/dielectric heterostructure to be formed in a single contact-and-release process without sacrificial-layers, their wet removal, or polymer surface treatments. As schematically illustrated in Fig. 1c, in this approach the interface of interest (MoS$_2$/SiO$_2$) is fabricated by embedding the desired receiving substrate (SiO$_2$) in a matrix (gold) that is chosen to have strong adhesive interactions with the material being transferred (MoS$_2$). When the hybrid substrate is brought into contact with the 2D material, the adhesive interactions with the matrix promote the transfer, despite the weak vdW interactions at the SiO$_2$ site. With this approach, the limits of vdW forces are overcome to enable ubiquitous vdW integration. Fig. 1f shows an example where a continuous monolayer of MoS$_2$ is cleaved directly from its source crystal on to a SiO$_2$-embedded-in-gold substrate. The observed 18.5 cm$^{-1}$ separation between the $E^1_{2g}$ and $A_{1g}$ Raman modes[24] (Fig. 1f) and the bright photoluminescence at 1.81 eV[25] (Supplementary Fig. 1) confirm presence of continuous monolayer-MoS$_2$ on SiO$_2$. This process is also scalable with large-area fabrication yields in excess of 80% measured for 2 μm features over millimeter-scale areas (Supplementary Fig. 1, Supplementary Note 1). This fabrication approach can be readily extended to other two-dimensional materials. Using the same adhesive matrix substrate, we demonstrate monolayer WSe$_2$ (Fig. 1g) and PtS$_2$ (Fig. 1h) directly transferred onto SiO$_2$ as evidenced by the absence of Raman peaks at 308 cm$^{-1}$ and 310-350 cm$^{-1}$ respectively[26,27]. Beyond these monolayer transition metal dichalcogenides, we show that few-layer post-transition metal monochalcogenide GaS can also be directly exfoliated forming vdW heterostructures (Fig. 1i)[28].

**Substrate engineering for adhesive matrix transfer**

Preparation of the receiving substrate through a template-stripping procedure is core to adhesive matrix transfer. In template-stripping[29], the structures of interest are fabricated on a carrier substrate, and subsequently peeled to reveal a planar surface which assumes the carrier's roughness. This allows fabrication of ultrasmooth (< 1 nm root-mean-square roughness) surfaces which enable conformal contact between the 2D material and the adhesive matrix, necessary to realize strong adhesive interactions[30]. This is evident in Fig. 2a,b where a ~1.7 nm rough thermally evaporated gold film is reduced in roughness to ~0.3 nm once template-stripped. Template-stripping also planarizes the feature of interest with the surrounding matrix. Beyond conformal contact, this allows the 2D material to uniformly span both surfaces without significant strain. Fig. 2c shows an atomic force microscope (AFM) image of a hybrid, template-stripped $SiO_2$/Au surface. Despite a slight bowing of the $SiO_2$ feature, no significant step-height is observed between the $SiO_2$ and the surrounding gold.

Rational selection of a matrix material with high adhesive forces is also necessary. In the materials system considered here (vdW semiconductor, bulk insulator, metal), we expect adhesive forces to be predominantly vdW interactions, yet, as discussed in the next section, depending on the choice of the adhesive matrix, forces beyond vdW can also be leveraged. The relative strength of the vdW forces can be understood from the Lifshitz theory[4]. Fig. 2d presents an extended Lifshitz model[31] describing the vdW interaction of monolayer-$MoS_2$ with itself, in addition to gold and $SiO_2$, discussed further in Supplementary Note 2. This analysis, the results of which agree with density functional theory (DFT) calculations[30,32], confirms the strong vdW interaction between $MoS_2$ and gold relative to that between $MoS_2$ and itself, illustrating gold's suitability as an adhesive matrix. The relatively weak interaction between $MoS_2$ and $SiO_2$, on the other hand, explains why direct exfoliation to this substrate is not possible.

Though this extended Lifshitz model explains the trends observed in experiments, a simplified model would further expedite rational materials selection. As discussed in Supplementary Note 2, the materials dependence of the vdW interactions can be captured by the Hamaker constant[4]. For a simple model of a material interacting with itself, this constant can be estimated using only a few parameters:[4,33,34]

$$A = \begin{cases} \dfrac{3}{16\sqrt{2}} \hbar\omega_b \dfrac{(n^2-1)^2}{(n^2+1)^{3/2}} & \text{Band Insulator} \\ \dfrac{3}{16\sqrt{2}} \hbar\omega_p & \text{Drude metal} \\ \dfrac{3}{4\pi} \hbar\omega_d & \text{Graphene} \end{cases} \quad (1)$$

where $\hbar$ is the reduced Plank's constant, $n$ is the refractive index of the insulator, $\omega_b$ is the band gap of the insulator, $\omega_p$ is the plasma frequency of the metal, and $\omega_d$ is the height of the Dirac cone in graphene. As discussed in Supplementary Note 2, the Hamaker constant between two dissimilar materials can then be approximated by the geometric mean of the self-interaction terms.

Despite its simplicity, this model, shown as square markers in Fig. 2d, agrees well with DFT calculations and experimental trends. The respectively large and small Hamaker constants of gold and $SiO_2$ relative to $MoS_2$ explain why transfer is possible to the former, but not the latter. This model further makes clear that vdW interactions depend on fundamental optical and electronic parameters ($\omega_b$, $n$, $\omega_p$, $\omega_d$). This inevitable dependence is what prevents the vdW forces at an interface from being decoupled from the properties of the constituent materials, hence, limiting fabrication of arbitrary functional interfaces through vdW forces alone. The evaluation of this model for several common materials is presented in Fig. 2e. It is evident that the direct transfer of $MoS_2$, hBN, $WSe_2$ or graphene to common insulators such as $SiO_2$, or $Al_2O_3$ is forbidden. This inability to integrate common 2D materials with bulk insulators is universal as discussed in Supplementary Note 2 and Supplementary Fig. 2. Our adhesive matrix transfer, however, allows such a forbidden transfer while ensuring it is scalable, pristine and versatile. To emphasize this versatility, we additionally demonstrate high-yield fabrication of $MoS_2$ on $Al_2O_3$ in Supplementary Fig. 3.

**Patterned transfer with a polymeric adhesive matrix**

Beyond clean, direct transfer of a pristine monolayer from its source crystal, the adhesive matrix transfer method can also accommodate aligned transfer of patterned materials, a feature not readily feasible conventionally but needed for the fabrication of functional structures and devices. We demonstrate this using lithographically patterned chemical vapor deposited (CVD) graphene on $SiO_2$. We chose CVD graphene for this demonstration for its well-established growth and patterning, despite the need for solution processing. Our technique is, however, versatile and can be readily extended to other source 2D materials, including those patterned in a dry manner[19,35,36].

An optical micrograph of patterned graphene as-fabricated on $SiO_2$ is shown in Fig. 3a. We use $SiO_2$ here as it is suitable for lithographic processing of graphene while making it visible under optical microscopy[37] to aid in alignment. Despite these advantages, the rigidity of $SiO_2$ prevents conformal contact between the graphene and template-stripped gold which as discussed in Supplementary Note 3 and Supplementary Fig. 4, prevents a continuous transfer. Thus, an alternative matrix material is needed.

With mechanical compliance and tunable chemical composition, polymers can provide both strong surface interactions and conformal contact, to serve as suitable and versatile adhesive matrices. To identify a polymeric adhesive matrix suitable for graphene, we tested template-stripped polydimethylsiloxane (PDMS), SU-8 and NOA-61 (Fig 3a). Candidate matrices were contacted with patterned graphene, and mildly heated to 65 ºC under ~1 MPa pressure in a nanoimprint tool to ensure conformal contact. Though no transfer is observed to PDMS, high-yield transfer is observed to both SU-8 and NOA-61, with feature sizes ranging from 2-100 μm. Transfer of arbitrary patterns is also achieved, with high-yields over millimeter-scale regions (Fig. 3b, Supplementary Fig. 5). We note that the transferred area is limited by our imprinting setup and can be extended with an improved apparatus. As discussed in Supplementary Note 4 and Supplementary Fig. 6, the adhesive character of these polymers is explained by their chemical composition and can consist of interactions beyond vdW. These polymers also work for diverse 2D materials, for example, CVD grown $MoS_2$ can also be dry-transferred to NOA-61, directly off of its growth substrate (Supplementary Fig. 5).

Raman spectroscopy of graphene transferred to NOA-61 (Fig. 3c) shows the characteristic $G$ and $G'$ ($2D$) peaks, in addition to negligible defect-associated $D$ peak, providing no evidence of damage during transfer[38]. Statistical characterization of the position and full-width-at-half-maximum (FWHM) of the $G$ and $G'$ peaks was further performed. Notably, the $G'$ peak shows a decrease in average FWHM from 53.8 to 32.0 cm$^{-1}$ after transfer (Fig. 3d), which suggests a relaxation in strain inhomogeneities[39]. The average compressive strain per Raman spot, as inferred from the positions of the $G$ and $G'$ peaks[40–42] (Supplementary Fig. 7), was further reduced from 0.045% to 0.024% after transfer, with a 2.5-fold reduction in the standard-deviation (Fig. 3e). Such strain relaxation is enabled by the deformability of the receiving polymer substrate, and can enhance transfer success by allowing the graphene to reduce its stored elastic energy[43]. Raman mapping and AFM imaging (Fig. 3f,g) further verify that the transferred material is continuous.

**Patterned, aligned fabrication of vdW heterostructures**

With a suitable adhesive matrix identified, we then demonstrate the alignment capability of our approach by forming arrays of graphene/gold heterostructures. First, the substrate was prepared by template-stripping lithographically-patterned gold features using NOA-61 as the adhesive matrix. The gold-embedded-in-polymer substrate was then aligned and contacted with patterned graphene-on-$SiO_2$. The substrates were exposed to mild heating and pressure, then separated, resulting in the transfer of graphene and the fabrication of aligned graphene/gold vdW heterostructures. A microscope image of the resulting substrate is shown in Fig. 4a, and a single series of 2-5 μm heterostructures is magnified in Fig. 4b. Raman mapping of the $G'$ peak shows that the transferred graphene is continuous (Fig. 4b). The Raman spectrum on gold shows no $D$ peak (Fig. 4c), indicating absence of any notable damage to graphene during transfer.

We further fabricated aligned, suspended graphene membranes. Such a structure represents the limit where no adhesive interactions are experienced between the 2D material and the receiving surface, and provides a platform for developing mechanically active devices. Given the fragility of monolayer-graphene, consistent with the past reports[44–47], we used double-layer (DL) graphene composed of two consecutively transferred monolayers. The DL-graphene was patterned, aligned and brought into contact with an NOA-61 substrate perforated with holes, then separated. The resulting aligned suspended membranes were imaged using differential interference contrast (DIC) microscopy, allowing for their direct visualization. Successful fabrication of a set of 2-4 μm suspended membranes is shown in Fig. 4d and supported by the Raman and AFM analyses in Fig. 4c,d.

Large-area analysis of alignment accuracy and fabrication yield was conducted via optical microscopy (Supplementary Note 1, Supplementary Fig. 8). For the graphene/gold features, we measured a root-mean-square alignment accuracy of 800 nm over a centimeter-scale chip (Fig. 4f). This already high alignment accuracy is limited by our manual alignment procedure, and thus does not represent the ultimate limit of our approach. Yield analysis was further conducted for both the graphene/gold and suspended DL-graphene samples, showing maximum yields of 85% and 80%, respectively (Fig. 4e). A decrease in fabrication yield was observed with an increase in feature size, consistent with the $MoS_2$/$SiO_2$ sample (Supplementary Fig. 1). As discussed in Supplementary Note 5 and Supplementary Fig. 9, we suspect that this size-dependent yield is due to defects introduced during materials growth or initial processing on $SiO_2$. Thus, we expect that the yields can be enhanced further by improving the material quality and processing to ensure minimized mechanical defects.

**2D material-to-device integration with pristine surfaces and interfaces**

By overcoming the limits of vdW interactions, adhesive matrix transfer can enable a single-step 2D material-to-device integration. In this strategy, there is no post-transfer fabrication, and no exposure to solvents, sacrificial layers, or high temperatures is necessary. Surfaces and interfaces remain pristine and intrinsic material properties are thus preserved. We demonstrate these features by fabricating arrays of monolayer-$MoS_2$ transistors. A representative optical micrograph of a fabricated device and its schematic illustration are shown in Fig. 5a,b. To make the transistors, the device elements are first fabricated and template-stripped to form the transfer substrate (Supplementary Fig. 10). Here, the gold surfaces serve as both the source/drain contacts and the adhesive matrix which enables the $MoS_2$ transfer across the $SiO_2$ channel. Once brought into contact with a crystal of $MoS_2$, a monolayer transfers and forms arrays of transistors, confirmed through optical microscopy and Raman spectroscopy (Supplementary Fig. 11).

Electrical characterization of the device in Fig. 5a reveals an on-off ratio of $\sim 10^8$ (Fig. 5c) and an on-state current of $\sim 3.7$ µA/µm (Fig. 5d, Supplementary Table 1). Ohmic contact behavior is revealed by the linear output characteristic for small drain-source voltages (Fig. 5d inset). An array of devices is characterized in Supplementary Fig. 11 and Supplementary Table 2. Notably, we observe n-type behavior in all measured devices. Previously, it has been shown that gold contacts transfer printed to multilayer-$MoS_2$-on-PMMA can display p-type behavior[15], however, for monolayer-$MoS_2$ and multilayer-$MoS_2$-on-$SiO_2$, only n-type behavior was observed[15,18,19]. Our approach allows the first demonstration of a transistor where monolayer-$MoS_2$ is directly transferred onto an electrode without any exposure to solvents or polymers, and is in a configuration where the 2D material is not sandwiched between the contact and a support substrate. This allows us to probe the pristine monolayer-$MoS_2$/gold interface, confirming the intrinsic n-type contact behavior in the absence of a support substrate or process contamination.

Because the top surface of the 2D material is kept clean and exposed, it can further be leveraged to tailor device performance (Fig. 5e). In particular, unique to our design, the surfaces above the 2D/metal contacts are available for further engineering. Leveraging this feature, here, we show that through charge-transfer doping with gold chloride ($AuCl_3$)[48], the fabricated n-type transistors can be engineered to give a p-type behavior. Though such p-type behavior is necessary for complementary logic applications, p-type contact to monolayer semiconductors has proven challenging[49]. Devices fabricated through our approach, as shown in Fig. 5g, display pure p-type behavior with no n-type branch and a high on-off ratio of $>10^4$, among the highest reported for any contact material (Supplementary Table 4). The consistent p-type performance (Supplementary Fig. 11, Supplementary Table 3) is enabled by the unique ability of our approach to fabricate clean vdW interfaces, with suppressed Fermi level pinning[15], while keeping the top surface accessible for efficient charge-transfer doping at the contact.

Beyond providing a platform to probe the intrinsic properties of 2D materials and their vdW heterostructures without the influence of processing damage, our approach can also allow for facile fabrication of devices with unconventional form-factors, such as those flexible (Fig. 5f). In contrast to existing approaches which require the devices to be directly fabricated on the final flexible substrate[50–52], in our approach the device features are first formed on a rigid carrier, then bonded and transferred to the final substrate (Supplementary Fig. 10,12). This eliminates the need for process compatibility between the substrate and the electrode fabrication procedure, allowing for the use of diverse substrate materials beyond the commonly used polyimide films[50–52]. Fig. 5h shows electrical characterization of an example flexible transistor on a polyethylene terephthalate (PET) backing, a substrate incompatible with conventional photolithography. This material versatility, combined with the pristine vdW integration offered by our adhesive matrix transfer, introduces new opportunities for flexible devices with 2D materials.

**Conclusions**

In this work, we developed the adhesive matrix transfer to extend vdW integration beyond the intrinsic limits of vdW forces to enable scalable, dry, and sacrificial-layer-free formation of diverse heterostructures and single-step 2D material-to-device integration. This is achieved by using an adhesive matrix to decouple the weak vdW forces forming the heterostructure of interest from the forces that promote its formation. With this approach, we demonstrated conventionally-forbidden 2D material-dielectric heterostructures, which are critical building blocks for functional devices. In this demonstration, the gold adhesive matrix surface interactions were largely dominated by vdW forces. However, through the choice of the matrix material forces beyond vdW can be leveraged to enable transfer of broader classes of materials as needed, without influencing the vdW nature of the desired heterostructure. As an example, using a polymeric matrix, we demonstrated scalable, dry and aligned transfer of CVD graphene to form patterned graphene/gold heterostructures and suspended membranes. Finally, we highlighted the unique features of adhesive matrix transfer for single-step material-to-device integration. Through fabricating arrays of monolayer-$MoS_2$ transistors with clean vdW interfaces, we were able to probe their intrinsic properties, and engineer them to realize both n- and p-type behaviors. Our transfer substrate preparation procedure is further compatible with arbitrary substrate materials, introducing a new platform for flexible devices. The adhesive matrix transfer is not limited to the material systems and designs discussed here. Through appropriate adhesive matrix engineering, this platform can be extended to other nanomaterials and applications where clean, dry, and large-area vdW integration is desired.

**Acknowledgements**

This work was supported by the National Science Foundation (NSF) Award CMMI-2135846, and NSF Center for Energy Efficient Electronics Science ($E^3S$) Award ECCS-0939514. P.F.S. and P.J.P. acknowledge support from the NSF Graduate Research Fellowship Program under Grant No. 1745302. Work by H.G., H.K. and X.L. was supported by the U.S. Department of Energy (DOE), Office of Science, Basic Energy Sciences (BES) under Award DE-SC0021064. H.G. acknowledges the support from Bunano Cross-Disciplinary Fellowship at Boston University. A.Y.L. and J.K. acknowledge support from the US Army Research Office through the Institute for Soldier Nanotechnologies at MIT, under cooperative agreement no. W911NF-18-2-0048. P.F.S. would like to thank the research staff of MIT.nano for supporting this work, in particular K. Broderick, W. Hess, J. Scholvin, D. Terry and D. Ward. P.F.S. would further like to thank J. Zhu for electrical measurement advice, N. Romeo for help with mechanical modelling, and M. Saravanapavanantham for characterization support.

**Author contributions**

P.F.S. and F.N. conceived of the project, designed the experiments, and wrote the manuscript. P.F.S. performed the experiments, and analyzed the results. P.F.S., P.J.P., W.Z. and F.N. developed the template-stripping process. W.Z. assisted with the PL measurements. P.F.S., M.T. and F.N. developed the $MoS_2$ transfer process. H.G. and X.L. grew the CVD $MoS_2$. S.Y.T, Y.L.C, C.N.K and C.S.L grew the $PtS_2$. Q.T. and X.L. grew the GaS. P.F.S, H.K., A.Y.L., J.K., X.L. and F.N. developed the graphene transfer process. F.N. supervised the project. All authors contributed to finalizing the manuscript.

**Competing interests**

The authors declare no competing interests.

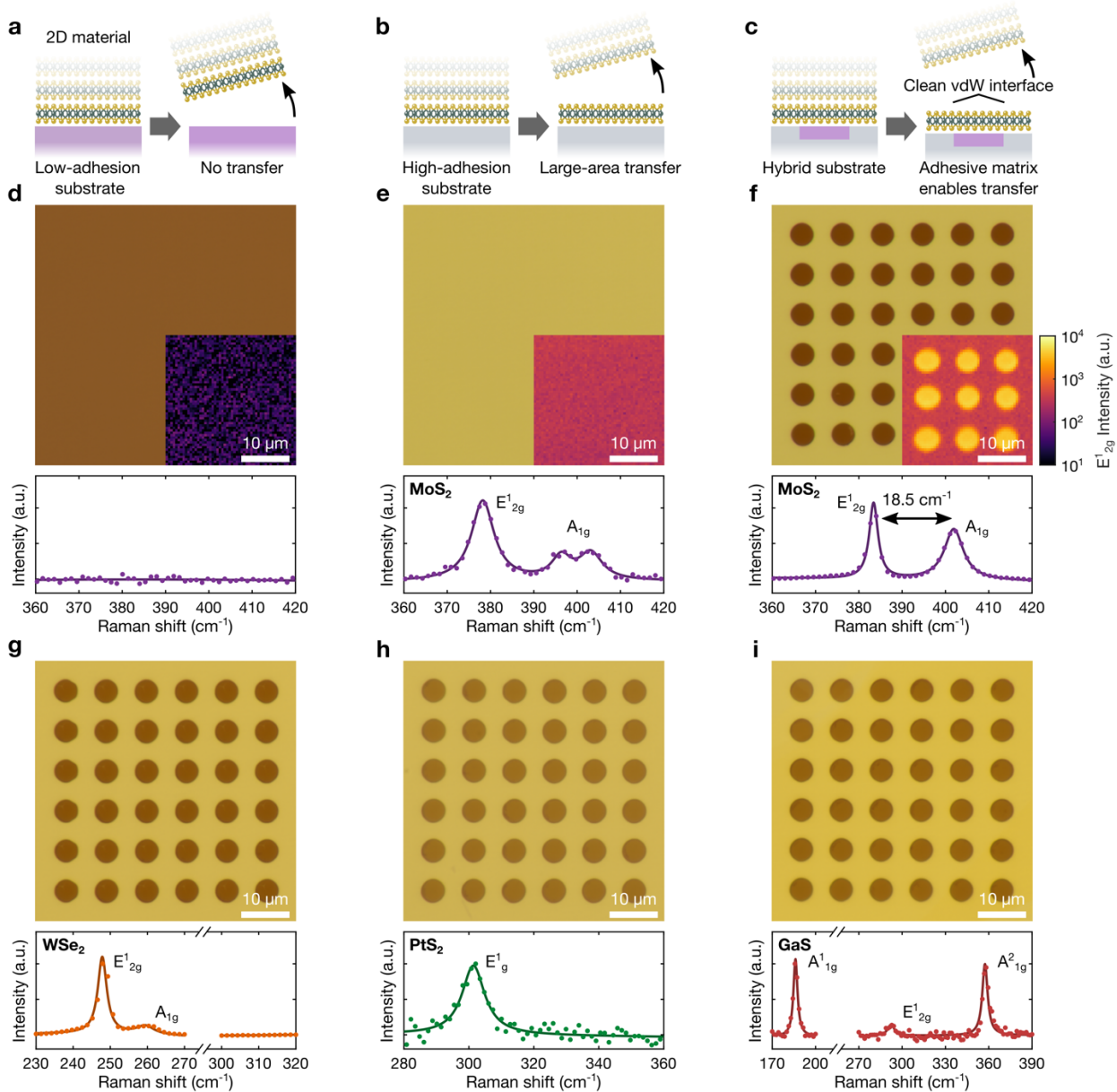

**Fig. 1 | Adhesive matrix transfer of pristine two-dimensional materials. a-c**, Schematic illustration of adhesive matrix transfer. **a**, When 2D material is brought into contact with a low-adhesion substrate no transfer is observed. **b**, Large-area transfer of continuous monolayer can be achieved with certain high-adhesion substrates. **c**, By embedding the low-adhesion substrate of interest in an adhesive matrix of high-adhesion material, clean vdW interfaces can be directly fabricated. **d,** Attempted transfer of MoS$_2$ directly to SiO$_2$. When samples are brought into contact and separated, no transfer is observed under optical microscopy (main panel), or Raman mapping of $E_{2G}^1$ peak integrated intensity (inset, same scale). Bottom panel of **d-i** show characteristic Raman spectra. **e,** When MoS$_2$ is brought into contact with template-stripped gold, continuous transfer over large areas is observed under optical microscopy and Raman mapping. Characteristic $E_{2G}^1$ and $A_{1g}$ peaks are observed on gold. **f**, By embedding low-adhesion SiO$_2$ in gold adhesive matrix, MoS$_2$/SiO$_2$ heterostructures can be directly fabricated, evidenced by Raman spectroscopy. Adhesive matrix transfer is readily extended for clean exfoliation of other 2D materials, here

transfer of monolayer transition metal dichalcogenides **g,** Wse$_2$, **h,** PtS$_2$, and of few-layer post-transition metal monochalcogenide **i,** GaS is demonstrated.

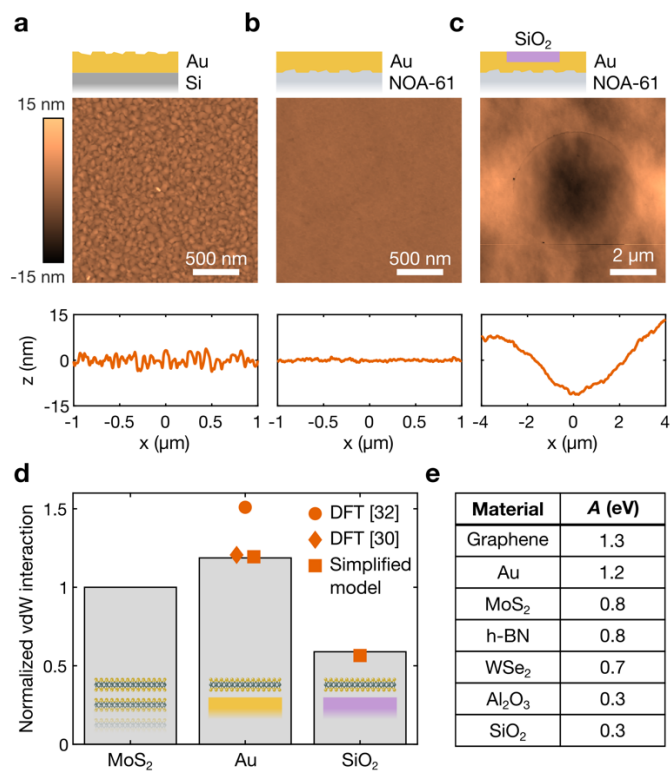

**Fig. 2 | Adhesive matrix engineering.** AFM characterization of 85 nm thermally evaporated gold film, **a**, as-deposited and **b**, after template-stripping showing reduction in surface roughness from ~1.7 nm to ~0.3 nm. **c**, AFM characterization of template-stripped SiO$_2$ embedded in gold. **d,** Extended Lifshitz model of MoS$_2$ interaction with itself, gold and SiO$_2$, and comparison with DFT (circle, diamond markers)[30,32] and simplified model (square markers). **e,** Simplified Lifshitz model of vdW self-interaction in several common materials showing that direct exfoliation of important 2D materials (graphene, MoS$_2$, WSe$_2$, hBN) to common dielectrics (Al$_2$O$_3$, SiO$_2$) is not possible. The model is discussed in detail in Supplementary Note 2, Supplementary Fig. 2 and Supplementary Table 5.

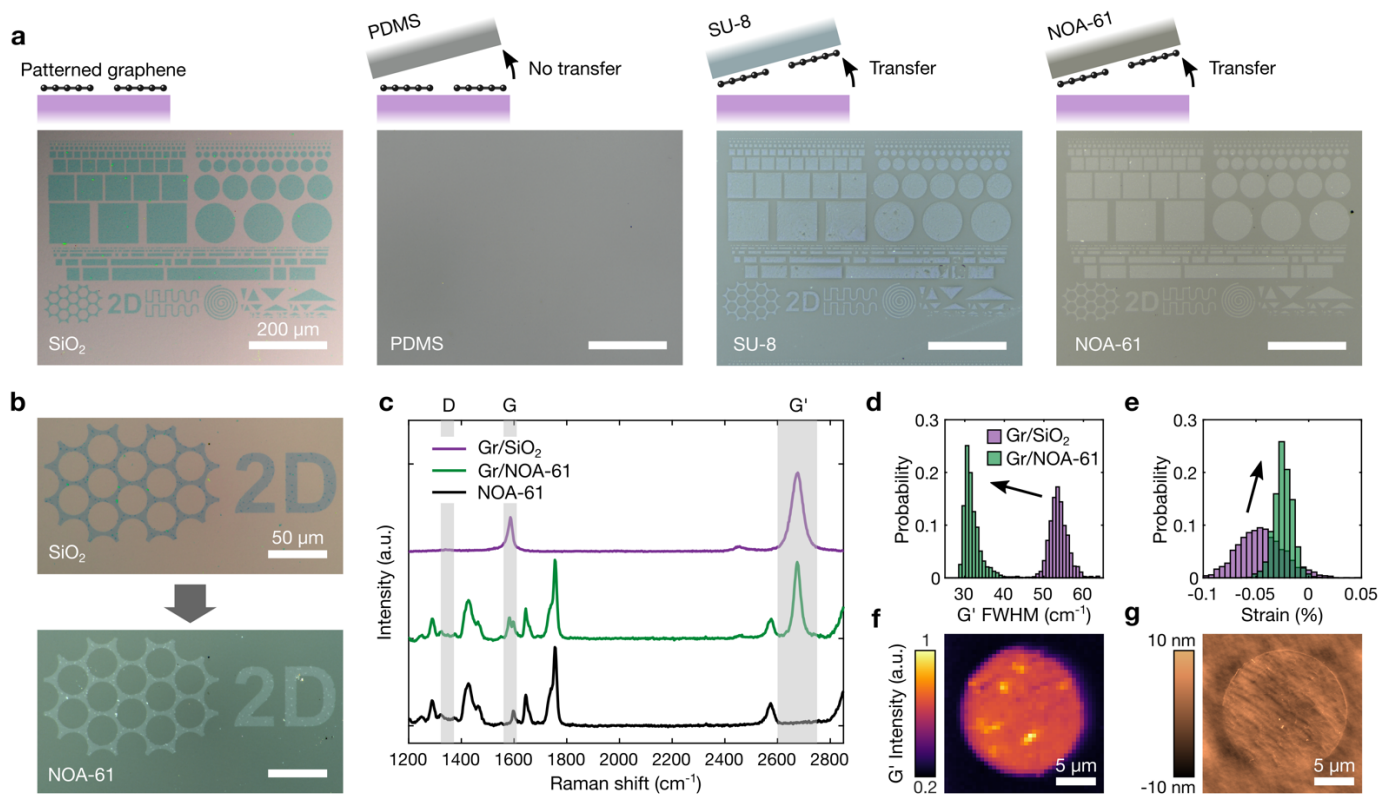

**Fig. 3 | Polymers as adhesive matrices. a**, Experiment for determining appropriate adhesive matrix for patterned CVD graphene. Patterned graphene on $SiO_2$ is prepared (left optical micrograph), then, brought into contact with different template-stripped polymers (right three optical micrographs). No transfer is observed to PDMS, however, high-yield transfer is observed to both SU-8 and NOA-61. **b**, Transfer of arbitrary patterns is achieved as evidenced by optical microscopy before and after transfer. **c**, Raman spectra show characteristic $G$, $G'$ peaks and minimal $D$ peak both for graphene (Gr) on $SiO_2$ and NOA-61. **d**, $G'$ peak shows reduction in full-width-at-half-maximum (FWHM) after transfer, characteristic of strain relaxation. **e**, Average strain, and strain distribution are reduced after transfer. **f**, Raman mapping of the integrated intensity of the $G'$ peak and **g,** AFM characterization of transferred graphene show continuous transfer.

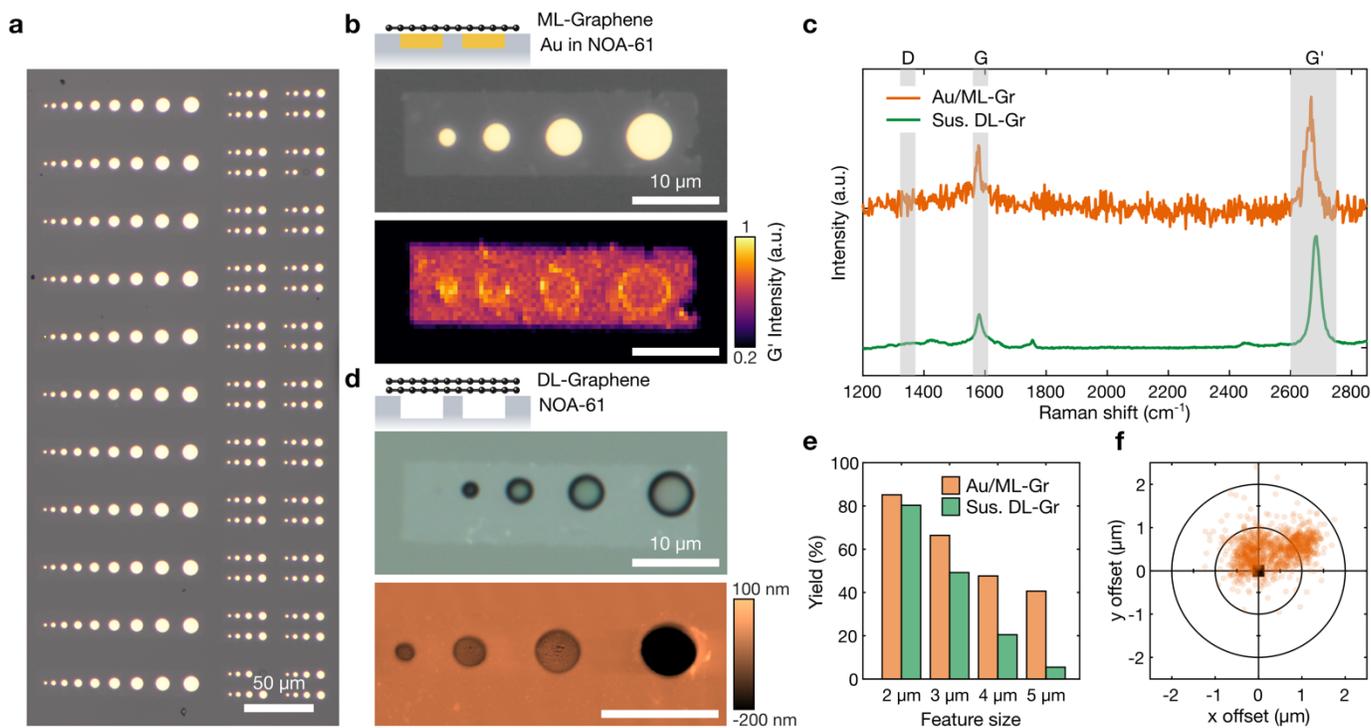

**Fig. 4 | Patterned, aligned fabrication of vdW heterostructures. a**, Large-area micrograph of patterned, aligned monolayer-graphene/gold heterostructures. **b**, Optical micrograph and Raman mapping of individual set of features, showing continuous monolayer (ML) graphene/gold interface across all feature sizes. Inset shows schematic illustration of the fabricated heterostructure. Raman map shows maximum intensity of $G'$ peak. **c,** Characteristic Raman spectra of ML-graphene on gold, and suspended double-layer (sus. DL) graphene. **d**, DIC and AFM micrographs of suspended DL-graphene. Successful suspension is observed for 2-4 µm features. **e**, Fabrication yield as a function of feature size for ML-graphene/gold heterostructures and suspended DL-graphene. **f**, Offset of $n = 1095$ features relative to desired position. Sub-micron alignment accuracy is demonstrated across 1 cm chip. Alignment and yield characterization are discussed in detail in Supplementary Note 1.

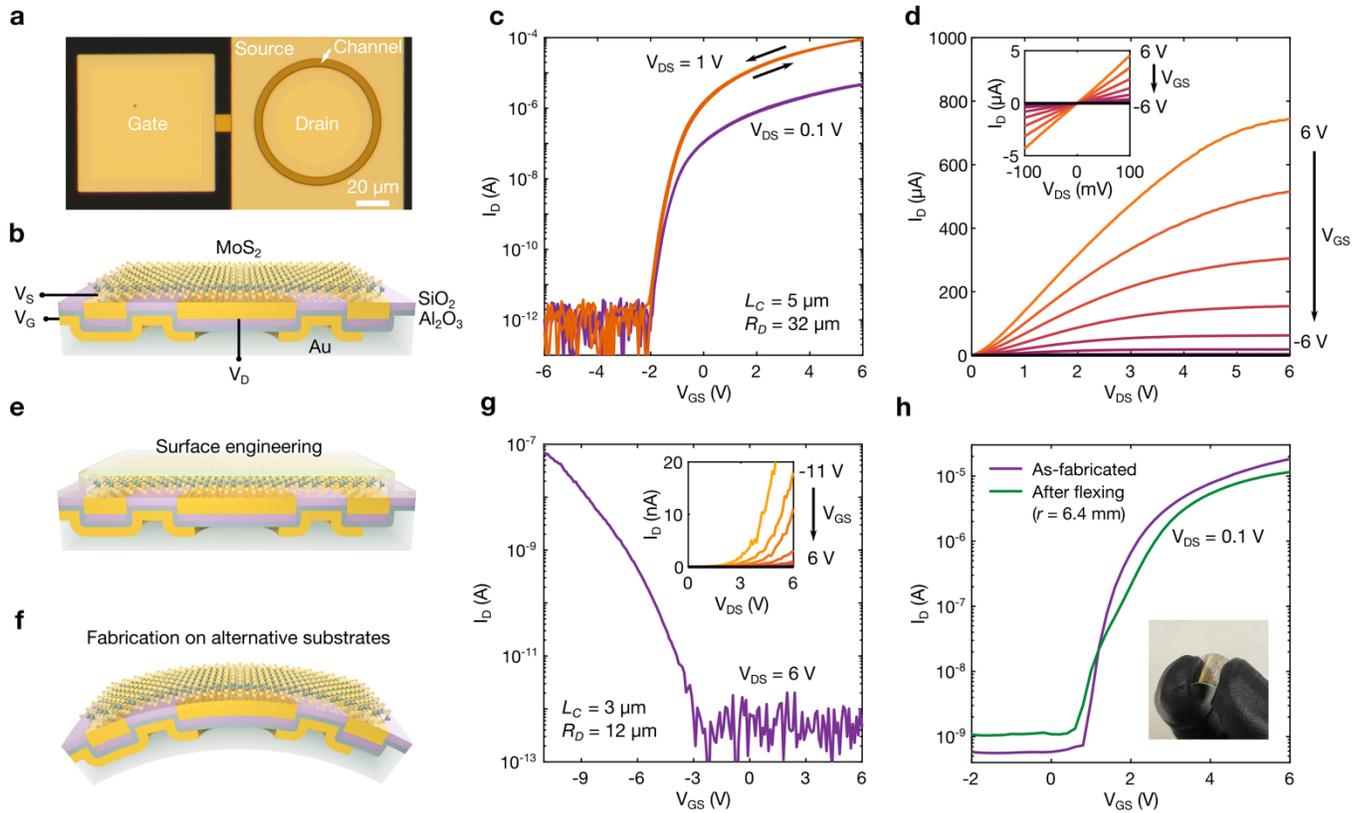

**Fig. 5 | Single-step adhesive matrix fabrication of pristine MoS₂ transistors. a**, Optical micrograph and **b,** schematic cross-section of a fabricated transistor. Here, channel length ($L_C$) is 5 μm and drain radius ($R_D$) is 32 μm **c,** Transfer characteristic (drain current, $I_D$ vs. gate-source voltage, $V_{GS}$) of transistor in **a**, showing high on/off ratio (~$10^8$), negligible hysteresis (< 100 mV) and good sub-threshold swing (202 mV/dec). **d**, Output characteristic ($I_D$ vs. drain-source voltage, $V_{DS}$) for $V_{GS}$ ranging from 6 V to –6 V in 1 V steps. Inset shows linear output characteristic for small $V_{DS}$, demonstrating ohmic contacts. **e,** Schematic illustration of surface engineering enabled by fabrication of a pristine device with exposed surface. **f**, Schematic illustration of device fabricated on flexible substrate, enabled by our unique electrode fabrication approach. **g**, Electrical characterization of p-type monolayer-MoS₂ device realized through AuCl₃ charge-transfer doping. Main plot shows transfer characteristic with high on-off ratio (>$10^4$) and unipolar p-type behavior. Inset shows non-linear output characteristic due to presence of Schottky barrier for holes. **h,** Transfer characteristic of device on PET substrate before and after flexing to radius of 6.4 mm. Inset shows photograph of fabricated flexible devices. Substrate is ~9 mm in size.

17